\documentclass[a4paper,11pt]{article}
\usepackage{graphicx, color}
\usepackage{natbib}
\usepackage{amsmath, amssymb}
\usepackage[hmargin=3cm]{geometry}

\usepackage{lineno}
\author{Nicolas \textsc{Brantut}}
\date{Department of Earth Sciences, University College London, London, UK\\
\textit{now at} GFZ Helmholtz Centre for Geosciences, Potsdam, Germany}

\title{Analysis of stress in the cohesive zone, dissipation and fracture energy during shear rupture experiments}

\begin{document}

\maketitle

\paragraph{Key Points:}\begin{itemize}
\item Stress change in cohesive zone can be computed from laboratory rupture experiments
\item Stress in cohesive zone is complex for slow ruptures
\item Rupture energy budget is consistent with dynamics, despite complexity in slip rate
\end{itemize}

\begin{abstract}
  We analyse high resolution slip rate data obtained during dynamic shear rupture experiments by \citet{berman20}. We use an inverse method to extract the details of strength evolution within the cohesive zone. The overall behaviour is slip-weakening at high rupture speeds ($>0.76C_\mathrm{R}$, where $C_\mathrm{R}$ is the Rayleigh wavespeed), but non-monotonic at low rupture speeds ($<0.76C_\mathrm{R}$), with a transient increase after an initial strong weakening. The \emph{slower} ruptures are associated to \emph{more} weakening in the cohesive zone. The fraction of breakdown work associated to the initial weakening, immediately behind the rupture tip, matches the fracture energy estimated by independent methods, but the total breakdown work can be much larger than fracture energy. Complex stress evolution in the cohesive zone is compatible with a well-defined fracture energy that explains rupture tip propagation, but the complexity is reflected in local slip rates that will impact radiated waves.

  %The emergence of a well-defined fracture energy that matches the rupture tip stress field is shown to be compatible with possibly complex cohesive zone stress profiles. However, the slip rate evolution is very sensitive to the details of the cohesive zone
\end{abstract}

\paragraph{Plain language summary} Ground motion during earthquakes is determined by the dynamics of fault slip at the earthquake source. An attractive approach to understand how faults slip and eventually generate seismic waves is to consider tectonic faults as thin fractures that propagate in an elastic material, so that existing knowledge on engineering fracture mechanics can be applied. This approach can be tested in the laboratory, and is typically shown to be successful. Here, we analyse recent laboratory data that show interesting departures from classical theory, and specifically determine the details of stress evolution during the slip propagation process. This analysis reveals that in some circumstances (here, for slow ruptures) the stress evolution is more complex than anticipated, which explains why the observed slip rate is markedly different from classical predictions. We show that this complexity is not necessarily incompatible with other predictions from fracture mechanics, notably in terms of energy balance.

\section{Introduction}

Can we use the theory of linear elastic fracture mechanics (LEFM) to describe earthquakes? This question has received considerable attention over the past decades, not least because LEFM provides tools to reduce the complex problem of earthquake propagation and arrest to that of an (hopefully simpler) energy balance problem that yields a so-called ``rupture tip equation of motion'' \citep[e.g.][]{husseini77}. Historically, earthquake sources have been modelled using LEFM \citep[e.g.][among many others]{kostrov66,madariaga77,freund79}, whereby the sliding region of an expanding rupture is assumed to sustain a uniform stress, and the potential nonlinearities occurring at the rupture tip where stress drops from a possibly finite peak strength to a constant residual are lumped into a scalar quantity called fracture energy. This description is only valid if the small-scale yielding hypothesis is satisfied, i.e., if the nonlinear region, also called the cohesive zone, is small compared to the total rupture size, and if no further changes in strength occur beyond the cohesive zone. Models that include a complete description of the cohesive zone are formally equivalent to LEFM in this case \citep[e.g.][]{rice68b,ida72,palmer73}.

However, it is not obvious \textit{a priori} that the small-scale yielding hypothesis is correct for earthquakes \citep[e.g.][]{kammer23}, or more generally for shear ruptures in materials. Indeed, the details of the strength degradation along faults are challenging to access. Only limited information can be accessed from seismological records, because of sparse data coverage, limited frequency bands and physical trade-offs between dynamic quantities \citep[e.g.][]{olsen97,peyrat01,tinti05,ruiz11}. Laboratory experiments that reproduce rupture propagation with extensive instrumentation have a good potential to constrain dynamic stress changes during shear ruptures. It has been shown that LEFM is good model for the onset of slip at interfaces between elastic bodies \citep[e.g.][]{svetlizky19}, in the sense that it correctly predicts rupture tip stress fields \citep[e.g.][]{svetlizky14,kammer19} and the motion of the rupture tip \citep[e.g.][]{bendavid10,bayart16}. Despite this success, in most laboratory experiments the cohesive zone remains elusive: it is typically small (from mm to cm) and transient (duration of the order of microseconds), and can only be observed if stress measurements are made directly \emph{on} the fault, or extremely close to it (typically at distances much smaller than the cohesive zone size itself; see Supplementary Information, Section 1). Most experimental data are not able to resolve any cohesive zone detail because measurements are made at some distance from the fault plane \citep[e.g.][]{johnson76,ohnaka87,okubo84,svetlizky14}. A few recent experimental studies where some details of the stress evolution could be resolved \citep[e.g.][]{kammer19,rubino17,paglialunga22} show that the cohesive zone can be more complex than previously assumed.

 Experimental work by \citet{paglialunga22} showed that multistage weakening can lead to the existence of several fracture energy values, appropriate for different stages of the rupture process. Using full-field imaging techniques during dynamic rupture experiments, \citet{rubino22} showed that slip between elastic blocks occurs in bursts with highly variable slip rate and traction evolution, which is at odds with the crack-like ruptures typically expected in the LEFM approximation.  The slip rate evolution along the fault is what determines near field strong motion and overall earthquake source-time functions, and it is thus important to determine if and how slip rate variations arise during dynamic fault motion.

In a recent study, \citet{berman20} reported slip rate data obtained during spontaneous dynamic propagation of shear cracks along a preexisting interface in PMMA. In their original paper, both linear elastic fracture mechanics (LEFM) and a simple cohesive zone model (a monotonically decreasing strength behind the rupture tip) were used to interpret the data. Their main conclusion was that LEFM predictions are a good match to slip rate data measured in fast ruptures, $C_{\rm f}\geq0.8C_{\rm R}$, where $C_\mathrm{f}$ is the rupture speed and $C_\mathrm{R}$ is the Rayleigh wave speed, and that a simple regularisation of the LEFM tip singularity by the cohesive zone model
\begin{linenomath}\begin{equation}\label{eq:expmodel}
  \tau(x) = (\tau_\mathrm{p}-\tau_\mathrm{r})e^{-x/x_\mathrm{c}} + \tau_\mathrm{r}
\end{equation}\end{linenomath}
produces a good match to the entire slip rate evolution behind the tip. In Equation \eqref{eq:expmodel}, $\tau_\mathrm{p}$ refers to the peak strength and $\tau_\mathrm{r}$ denotes the residual strength of the interface, $x$ is the along-fault coordinate, and the quantity $x_\mathrm{c}$ is the characteristic size of the cohesive zone.

However, neither LEFM nor Equation \eqref{eq:expmodel} produced a satisfying match to slip rate data obtained in ``slow'' ruptures, $C_\mathrm{f}<0.8C_\mathrm{R}$, despite the overall good match of LEFM with ``far-field'' strain data. The complexity in slip rate evolution \citep[figure 4 of][]{berman20} clearly calls for an equally complex strength evolution within the cohesive zone.

Here, we use \citeauthor{berman20}'s high resolution slip rate data in an inverse problem to determine the cohesive strength. This approach allows us to systematically explore the features of the cohesive strength that produce the observed slip rate, and determine the key differences between slow and fast ruptures. In addition, we use our estimate of strength evolution to determine the energy dissipation (specifically, the breakdown work \citep{tinti05}), and compare it with independent estimates of fracture energy. Overall, we find that slow ruptures tend to have a complex stress evolution, which includes substantial strengthening. Despite that complexity, the breakdown work matches well with the elastically-inferred fracture energy. Thus, LEFM seems to be a good approximation in terms of energy balance, but cohesive zone complexity leads to clear differences in local slip rate evolution.

\section{Method}

%\subsection{Data}

The data used are the slip rate profiles and rupture speeds from \citet{berman20} determined by optical methods during dynamic ruptures running along a Polymethylmethacrylate (PMMA) interface. In their experiments, \citet{berman20} sheared a narrow (5.5~mm) slab of PMMA (length 150~mm) on top of a PMMA base, and determined rupture tip position and speed by optically imaging the real area of contact between the blocks. An array of strain gauges positioned 3.5~mm away from the surface was used to measure dynamic strains (and obtain fracture energy of each rupture), and a dedicated optical interferometer was developed to measure slip and slip rate on a small patch (5~mm in length) along the fault. Details of the experimental setup and measurement methods are given in the original work of \citet{berman20}. Note that \citet{berman20} report their results in terms of particle velocity $v_x$, which ought to be multiplied by 2 to obtain the slip rate.

% In the original paper, the authors mentioned that they compute $v_x$, the particle velocity, which is one half of the slip rate. However, the figure captions all mention $v_x$ as the slip rate. I assumed here that what they plotted was indeed the particle velocity, which must be multiplied by 2 to get the slip rate.

%\begin{figure}
%  \centering
%  \includegraphics{figures/sliprate}
%  \caption{Slip rate data from \citet{berman20}. The tip position is not the same as reported originally: it was adjusted for each curve by picking the data point where the slip rate starts to increase.}
%  \label{fig:sliprate}
%\end{figure}

%\subsection{Inversion}

The material surrounding the interface is assumed to be linear elastic. We are only interested in the near-tip region, and thus consider the approximation where the rupture is semi-infinite, i.e., the other rupture tip is far, driven by a negligible stress drop. We also follow the original analysis of \citet{berman20} and assume the rupture is locally at steady-state (constant rupture speed). Elastodynamic equilibrium implies a relationship between slip rate $V$ and shear stress $\tau$ in the rupturing patch \citep[e.g.,][]{viesca15}:
\begin{linenomath}\begin{equation} \label{eq:elastodyn}
  \tau(x) - \tau_\mathrm{b} = \frac{\bar{\mu}}{2\pi C_\mathrm{f}}\int_0^\infty\frac{V(\xi)}{\xi-x}d\xi,
\end{equation}\end{linenomath}
where $x$ is the position along the rupture ($x=0$ at the tip), $\tau_\mathrm{b}$ is the background stress, and $\bar{\mu}$ is a modified shear modulus that depends on the rupture speed $C_\mathrm{f}$. In mode II, we have \citep{rice05}
\begin{linenomath}\begin{equation}
  \bar{\mu} = \frac{\mu}{1-\nu}\times \frac{4\alpha_\mathrm{s}\alpha_\mathrm{d}-(1+\alpha_\mathrm{s}^2)^2}{\alpha_\mathrm{s}(1-\alpha_\mathrm{s}^2)},
\end{equation}\end{linenomath}
where $\alpha_\mathrm{s,d}=\sqrt{1-(C_\mathrm{f}/C_\mathrm{s,d})^2}$, with $C_\mathrm{s}$ and $C_\mathrm{d}$ the S and P wave speeds of the surrounding material.

For steady rupture, we also have the following relation between slip rate $V$ and slip $\delta$:
\begin{linenomath}\begin{equation}
  \delta(x) = \int_0^x V(\xi)/C_\mathrm{f}d\xi.
\end{equation}\end{linenomath}

It is tempting to use Equation \eqref{eq:elastodyn} directly, using the slip rate obtained in the experiments and computing the integral to obtain the shear stress. However, this strategy is impractical: The upper integration bound should extend to infinity, but the slip rate data only span a narrow region near the tip, so the integral cannot be computed unless we severely extrapolate the slip rate data. The alternative strategy used here is to determine $\tau(x)$ via an inverse method.

We assume $\tau(x)$ to be a piece-wise linear function, parameterised by its value $\tau_i$ at a set of positions $x_i$ ($i=0,\ldots,N$), valid for $x\in(0,+\infty)$:
\begin{linenomath}\begin{equation} \label{eq:tau}
  \tau(x) = \left\{
    \begin{array}{l}
      \tau_{i-1} + (\tau_{i}-\tau_{i-1})(x - x_{i-1})/(x_i - x_{i-1}) \quad \text{if } x \in [x_{i-1},x_i],\\
      \tau_N\quad \text{if } x>x_N,
    \end{array}\right.
\end{equation}\end{linenomath}
where $x_0 = 0$ and $x_N$ is the maximum position where slip rate was recorded. The shear stress $\tau_N$ is imposed at all positions beyond $x_N$, which means that $\tau_N$ is a constant residual stress. To ensure consistency with the semi-infinite crack approximation, the residual stress $\tau_N$ is imposed equal to the background stress $\tau_\mathrm{b}$. We fix the positions $x_i$ at 64 logarithmically-spaced points from the rupture tip to the maximum extent of the slip rate data. The unknowns of our problem are thus the values of $\tau_i$ at all positions $x_i$.

As our forward problem, we use the stress evolution \eqref{eq:tau} to compute the associated slip rate from \eqref{eq:elastodyn}, which can be done numerically. Formally, our direct problem is expressed as
\begin{linenomath}\begin{equation}
  \mathbf{d} = \mathbf{g}(\mathbf{m}),
\end{equation}\end{linenomath}
which allows us to compute data predictions $\mathbf{d}$ (a set of slip rate values at increasing positions from the crack tip), from a vector of model parameters $\mathbf{m}$ (the shear stresses $\tau_i$). The function $\mathbf{g}$ is the solution of \eqref{eq:elastodyn} for slip rate, knowing the shear stress evolution; it is computed numerically using Gauss-Chebyshev quadrature \citep{viesca18}.

%, and . In practice, we set up an inverse problem where we have a vector of observations $\mathbf{d}_\mathrm{obs}$, here our slip rates at different positions from the tip, and a physical model

The inverse problem is solved by the quasi-Newton method \citep{tarantola05}, using automatic differentiation to compute the Jacobian of $\mathbf{g}$ \citep{revels16}. The details of the inversion procedure are given in the Supplementary Information, Section 2. For each slip rate profile, we obtain a mean model (in the least square sense) for the best-fitting shear stress profile. Synthetic tests show that the results are not sensitive to the detailed choice of rupture tip position, and that the inverted shear stress profile is only constrained up to a constant, uniform background (see Supplementary Information, Section 3). 

\section{Results}

\begin{figure}
  \centering
  \includegraphics{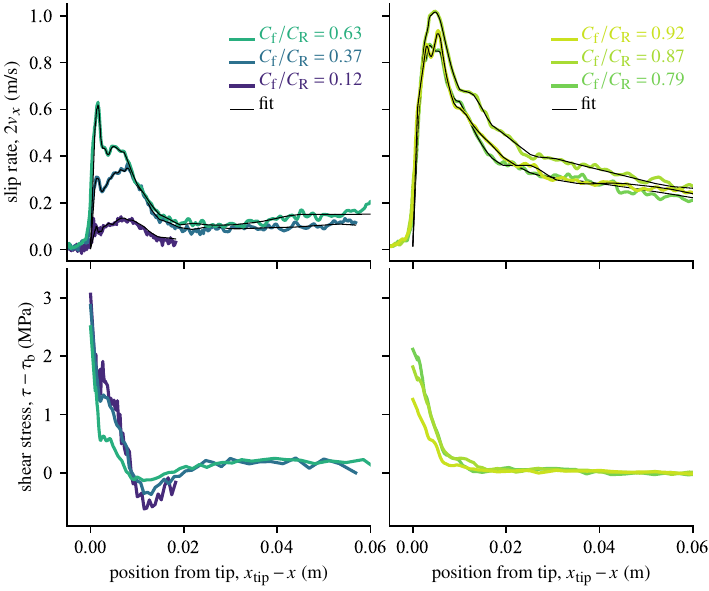}
  \caption{Examples of shear stress evolution inverted from slip rate profiles.}
  \label{fig:examples}
\end{figure}

Representative examples of slip rate fits and corresponding shear stress are given in Figure \ref{fig:examples}. As originally reported by \citet{berman20}, there is a clear difference between slow ($C_\mathrm{f}<0.76C_\mathrm{R}$) and fast cracks: slow ruptures tend to be associated with nonlinear and nonmonotonic traction evolution. The two slip rate peaks occurring during slow ruptures appear to be linked to a two-stage weakening, with an initial rapid stress drop, followed by a slower decay. The increase in slip rate at some distance from the tip is explained by a stress rebound (strengthening). By contrast, fast ruptures are associated to very simple traction evolution, a monotonic, almost linear weakening behaviour with constant residual stress.

\begin{figure}
  \centering
  \includegraphics{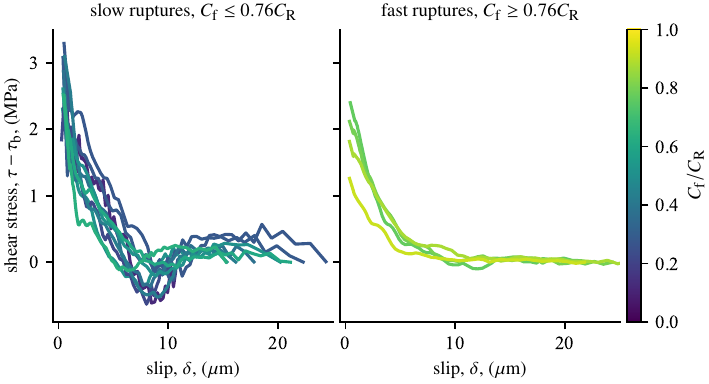}
  \caption{Shear stress evolution vs. slip obtained from inversion of slip rate profiles.}
  \label{fig:slipweakening}
\end{figure}

\begin{figure}
  \centering
  \includegraphics{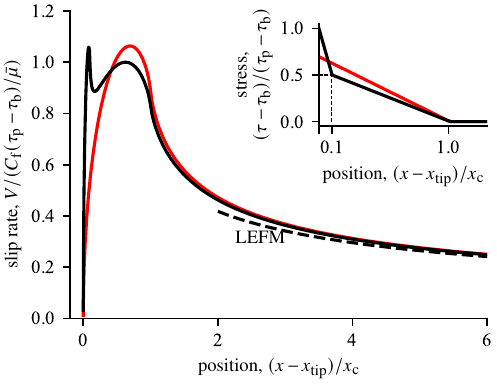}
  \caption{Typical slip rate profile obtained from two successive weakening stages in a semi-inifinite, dynamically progating crack solution (Equation \ref{eq:elastodyn}). The background stress is assumed equal to the residual strength. The strength decreases sharply from a peak value $\tau_\mathrm{p}$ down to $\tau_\mathrm{p}/2$ over a distance $x_\mathrm{c}/10$, and then decreases slowly down to the residual at a distance $x_\mathrm{c}$ from the tip (inset). The dashed line shows the slip rate obtained from the LEFM limit with a fracture energy consistent with the slip-weakening behaviour. For comparison, the slip rate associated to a single linear weakening stage (with the same fracture energy) is shown in red.}
  \label{fig:twoslopes}
\end{figure}

The shear stress profiles along the crack can be plotted as a function of the cumulated slip (Figure \ref{fig:slipweakening}). The behaviour is clearly different between slow and fast cracks, with slow ruptures associated with a clear restrengthening at around 8~\textmu m slip. In those slow ruptures, the shear stress peaks at about 3~MPa above the background stress, and then drops substantially below $\tau_\mathrm{b}$ (i.e., strong dynamic strength drop), prior to rapidly recovering. At large slip, the strength approaches a constant as the interface gradually restrengthens, but it is not certain that a constant residual is achieved fully. One important feature of the slip rate evolution, the existence of a first peak followed by a second, more gradual ``bump'', is specifically caused by the presence of two weakening stages: An initial rapid weakening, typically occurring over the first 1\textmu m of slip followed by a more gradual one, as can be confirmed by independent forward simulations (Figure \ref{fig:twoslopes}).

By contrast, consistently with the original work of \citet{berman20}, the slip rate during fast ruptures is simply explained by a monotonic decay of strength and stabilisation to a constant residual. The details of stress evolution revealed by the inversion method show that one of the fast ruptures (at $C_\mathrm{f}/C_\mathrm{R}=0.87$) also includes a short, fast initial weakening stage within the first 0.5\textmu m of slip (Figure \ref{fig:slipweakening}). This indicates that the two-stage weakening might persist even at high rupture (and slip) velocities, but might be less visible due to a vanishing difference between the two stages.

%All our inverted shear stress values appear to peak at around 1~MPa above the background shear stress, which is itself not accessbile from the inversion technique.

\begin{figure}
  \centering
  \includegraphics{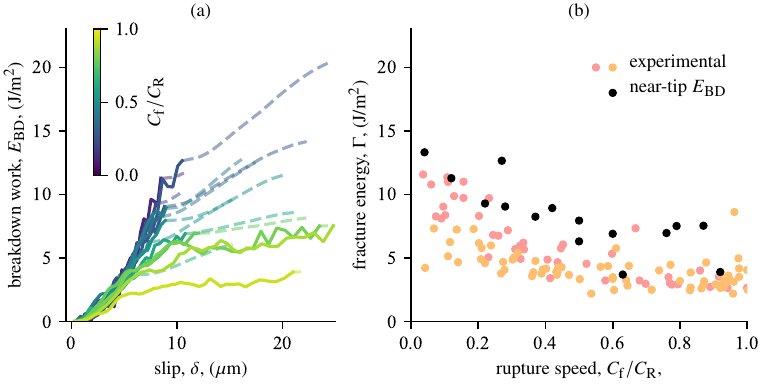}
  \caption{Energy dissipation (a) and estimate of fracture energy (b) from the inverted shear stress evolution. The dashed portions of curves in (a) correspond to the restrengthening phase of the stress-slip evolution. Black dots in (b) correspond to fracture energy estimated using the inverted stress-slip curves, pink and orange dots are independent estimates made by \citet{berman20} based on strain gauge measurements (fitting of stress variations during rupture assuming singular rupture). The two colors correspond to two experiments conducted in the same conditions.}
  \label{fig:energy}
\end{figure}

The access to shear stress evolution allows us to compute the energy dissipation within the cohesive zone of the dynamic ruptures. The quantity of interest is the so-called ``breakdown work'' \citep[e.g.][]{tinti05}. For nonmonotonic shear stress vs. slip behaviour, the breakdown work can be defined as
\begin{linenomath}\begin{equation} \label{eq:ebd}
  E_\mathrm{BD} = \int_0^\delta \tau(\delta') - \tau_\mathrm{min} d\delta',
\end{equation}\end{linenomath}
where $\tau_\mathrm{min}$ is the minimum stress reached in the interval $(0,\delta)$. The breakdown work is (by this definition) an increasing function of cumulated slip. We observe a clear stabilisation of breakdown work for fast ruptures (Figure \ref{fig:energy}a). For slow ruptures, the breakdown work reaches a plateau when restrengthening occurs (at around 8~\textmu m slip), and keeps increasing beyond that point: there is more dissipation away from the crack tip.

We can also use the breakdown work to estimate the fracture energy $\Gamma$. For slip-weakening cohesive laws with a well-defined residual, the fracture energy is simply the limit of $E_\mathrm{BD}$ at large slip \citep{palmer73}. For nonmonotonic cohesive law, we may identify $\Gamma$ with the ``near-tip'' dissipation, i.e., the fraction of $E_\mathrm{BD}$ that is associated with the initial weakening, down to the minimum stress achieved along the crack. The fracture energy estimated this way (Figure \ref{fig:energy}b) is of similar magnitude and decreases with increasing rupture speed in a similar way to that determined by \citet{berman20} based on strain gauge data fitted to a singular crack tip stress field (their Figure S4).

\section{Discussion and conclusions}

The present analysis brings quantitative constraints to the main result obtained by \citet{berman20}: slow ruptures are characterised by slip rate profiles that are markedly different from the predictions of LEFM and from simple weakening cohesive laws. Indeed, for ruptures where $C_\mathrm{f}<0.76C_\mathrm{R}$, the inversion procedure shows that shear stress initially decreases, reaches a minimum, but increases again. This observation is systematic, and markedly different from what happens in fast cracks. In addition, the presence of two separate slip rate peaks shows that weakening occurs in two stages, with an initial abrupt drop followed by a slower decrease.

The two-stage weakening is clearly manifested by the two peaks in slip rate observed in the slow ruptures, which is confirmed by simulations (Figure \ref{fig:twoslopes}). Such a two-stage weakening was already inferred from near-fault stress measurements in similar experiments by \citep{paglialunga22}, but its impact on slip rate in the cohesive zone is now clearly measured. The physics of weakening in PMMA is obviously different from that in rocks in natural fault zones, but multi-stage weakening is a likely possibility in natural earthquakes due to existence of a series of weakening processes, from flash heating, thermal pressurisation \citep[e.g.][]{noda09,viesca15}, thermal weakening \citep[e.g.][]{hirose05,harbord21}, coupled to fluid dilatancy and diffusion effects \citep{brantut21b}. As predicted by LEFM, the effect of two-stage weakening is restricted to the cohesive zone, and the expected classical solution (with a decay proportional to $1/\sqrt{x-x_\mathrm{tip}}$) should emerge at large distances to the tip (Figure \ref{fig:twoslopes}, dashed lines). In the dataset from \citeauthor{berman20}, the slip rate is markedly different from that limit due to a restrengthening effect.

The cause of strengthening in the cohesive zone at low rupture speed is not clear. PMMA has a very low melting temperature (only around 120~K above ambient temperature), and it is possible that local melting at asperity contacts occurred during the tests reported by \citet{berman20}. In rocks, the onset of melting during high velocity friction tests is often associated with a transient strengthening, leading to the so-called ``viscous break'' effect \citep{hirose05}; this process might have occurred in the PMMA experiments. Such nonmonotonic strength evolution was not observed in similar rupture experiments conducted in Homalite \citep{rubino17}, which displayed only monotonic weakening with ongoing slip that could be explained quantitatively by flash heating \citep{rice06}. By contrast, recent work on fault gouge by \citep{rubino22} also showed slip-strengthening behaviour at low slip rate. Thus, it is likely that the details of the cohesive zone exhibited here are material-dependent or microstructure-dependent. For natural earthquakes, strength recovery at some distance from the rupture front could occur due to late-stage melting (viscous break effect) or strength recovery due to the decrease in slip rate as would be predicted by flash heating \citep[e.g.][]{harbord21}. Such restrengthening may have important consequences for the dynamics of rupture and earthquake scaling laws \citep[e.g.][]{gabriel24}.

%In general, the behviour observed here with PMMA samples is consistent with 

\begin{figure}
  \centering
  \includegraphics{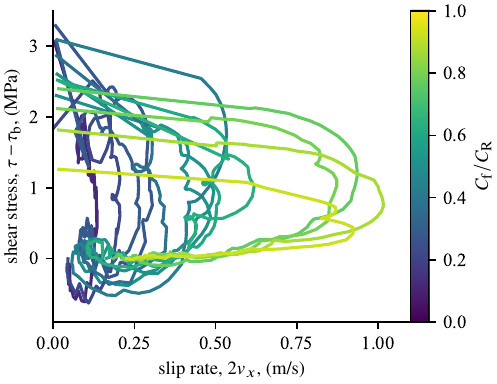}
  \caption{Cohesive zone stress evolution as a function of slip rate.}
  \label{fig:velocityweakening}
\end{figure}

One counterintuitive aspect of the strength evolution inverted from slip rate data is that \emph{more weakening} is observed during slow rupture compared to fast ruptures (Figure \ref{fig:slipweakening}), which translates into more weakening at low slip rate (Figure \ref{fig:velocityweakening}). The shear stress evolution within the cohesive zone can be interpreted as ``friction'', defined in the narrow sense as the constitutive relationship between strength, slip rate and possibly other (unknown) variables. In this framework, the behaviour observed here in PMMA samples can be interpreted qualitatively in terms of a competition between the so-called ``direct effect'' in rate and state friction \citep{dieterich79,ruina83}, which produces an instantaneous strengthening upon slip acceleration, and the evolution of one or several internal state variables that characterise the microstructure of the interface, which here could produce weakening. An analysis of shear crack propagation with rate-and-state friction as constitutive law in the cohesive zone, given by \citet{garagash21}, demonstrates that one expects increasing strength drop and larger fracture energy with increasing rupture speed, in contrast to the observations of Figures \ref{fig:energy} and \ref{fig:velocityweakening}. Therefore, the conventional rate-and-state framework (with a single state variable) is not \textit{a priori} consistent with the PMMA experiments. A few possibilities might be envisioned to explain the data. There might exist a true threshold in slip rate (and hence in rupture speed) above which the frictional behaviour of PMMA changes character, which could be consistent with flash heating or frictional melting (as reported originally by \citet{berman20}). This option is akin to modelling friction in the cohesive zone with two or more state variables, each having independent dependencies on slip and slip rate, possibly activated by temperature. There might also be an effect of prior state, whereby fast ruptures tend to occur along interfaces that experienced less contact healing (i.e., the initial strength is far from steady state). In this case, one would expect an effect of the particular rupture sequence in the experiments, which was not documented by \citet{berman20}.%The rapid acceleration observed during fast ruptures could lead to a more dominant direct effect, and thus reduce the total weakening within the cohesive zone. %At so-called ``intermediate'' velocity, i.e., around 1 to 10~mm/s, many materials exhibit highly variable frictional behaviour, as documented for instance in rotary-shear experiments (which provide a close approximation of the frictional rheology of a material point)

The details of the cohesive zone revealed by the inverse model illustrate a crucial phenomenon in fracture mechanics that could be overlooked if we focus only on the local friction law of the material: more weakening in the cohesive zone does not necessarily imply faster rupture speeds. Here, the opposite is observed. This observation can be explained by considering that the fracture energy associated with slower events seems larger than, or at best of similar magnitude to, that associated to faster events (Figure \ref{fig:energy}). We thus observe directly the disconnect between the details of the weakening in the cohesive zone and fracture energy, which is an integrated quantity that drives rupture propagation and is insensitive to such details. The complex cohesive law is not in contradiction with the existence of a well-defined fracture energy and LEFM behaviour at scales much larger than that of the cohesive zone.

It is difficult to make a clear sense of the breakdown work as defined in \eqref{eq:ebd} beyond the minimum stress achieved in the cohesive zone (Figure \ref{fig:energy}(a), dashed lines): this energy keeps increasing with increasing slip, and we see a clear disconnect between the fracture energy that drives the rupture (edge-localised dissipation) and the overall dissipation in the interior of the growing rupture. What cannot be addressed with the existing dataset is whether the stress would continue to rise well beyond the tip region, or if more weakening occurs. The stress evolution away from the rupture tip contributes to the energy release rate and thus to the overall dynamics of rupture, especially during rupture arrest \citep[e.g.][]{paglialunga22}.

In conclusion, the complex stress evolution inferred during ``slow'' ruptures ($C_\mathrm{f}<0.76C_\mathrm{R}$) along PMMA interfaces highlights that there might be regimes where LEFM is only applicable in an ``effective'' sense: one can use LEFM concepts (fracture energy) to understand, to first order, the dynamics of rupture expansion, but using physical quantities that are not necessarily well-defined material parameters. In this sense, the fracture energy derived here for slow ruptures (Figure \ref{fig:energy}b) is the ``effective'' energy that would produce similar dynamics to that of an ideal LEFM rupture, but does not capture all aspects of that rupture. In particular, the slip rate evolution is critically dependent on details of the cohesive zone not captured by LEFM. This has implications for earthquakes: the slip rate history on the rupture plane (source-time function) is what determines ground shaking and the far-field radiation measured in seismograms, and it remains to be seen to which extent LEFM (or a modification thereof) can capture both earthquake propagation and radiation.

% Apart from %

% show figure of stress change vs. slip rate: there is something here that needs to be understood! More weakening at low slip rate/low rupture speed. How to make sense of that?

%Remarkably, the complex cohesive law is not contradicting the existence of a well-defined fracture energy and LEFM behaviour at scales much larger than that of the cohesive zone. 

%I suspect that the inversion procedure used here can be applied systematically to extract cohesive laws for dynamic ruptures, which could then be compared to ``friction laws'' (i.e., detailed models of strength evolution at very small slip distances and high speed).

\paragraph{Acknowledgments} I thank Fabian Barras, Neri Berman, Jay Fineberg, Federica Paglialunga and Fran\c cois Passel\`egue for key discussions and feedback during the (long) preparation of this manuscript. Comments by Dmitry Garagash and an anonymous reviewer helped clarify many points of the paper. Rob Viesca helped clarify some review comments. Funding from the UK Natural Environment Research Council (grant NE/S000852/1), the European Research Council under the European Union’s Horizon 2020 research and innovation programme (project RockDeaf, grant agreement 804685; project RockDeath, grant agreement 101088963), and a Philip Leverhulme Prize from the Leverhulme Trust, is gratefully acknowledged.

\paragraph{Open Research} No new data have been generated in this work. Existing data from \citet{berman20} were used.

%\bibliographystyle{agufull}
%\bibliography{localbib}

\end{document}

\maketitle

\section{On- and off-fault dynamic near-tip stresses}

In rock or polymer friction experiments, strain gauges are commonly used to measure the evolution of the stress field at given locations on fault walls. If a strain gauge is located ``very near'' the fault plane, we might be tempted to use the record as representative of the ``on-fault'' stress, i.e., the strength. However, this might be inaccurate: elastic stress transfer (including waves) tends to smear out the stress evolution far from the fault. How close from the fault do we have to position strain gauges to reliably access the cohesive zone properties?

Here, we use a simple cohesive zone model, for which closed-form solutions are readily available, and compare the resulting near-fault strain field to that of a singular rupture model of the same energy. We establish that strains have to be measured at distances much smaller than that of the cohesive zone length to resolve its details. In practice, this is hard to achieve experimentally.

\begin{figure}
  \centering
  \includegraphics{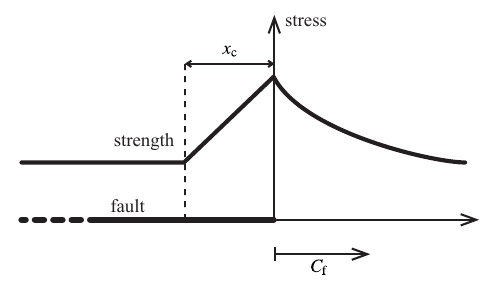}
  \caption{Rupture tip model, with a semi-infinite crack propagating at constant speed $C_\mathrm{f}$, and a linear weakening region of size $x_\mathrm{c}$. See \citet{poliakov02}.}
  \label{fig:sketch}
\end{figure}

Consider a rupture tip propagating at constant rupture speed $C_\mathrm{f}$. The fault strength is assumed to degrade with ongoing slip, in such a manner as to produce a linear strength evolution with increasing distance from the rupture tip (Figure \ref{fig:sketch}); this is the model analysed by \citet{poliakov02}. This model introduces a cohesive zone length $x_\mathrm{c}$. Asymptotically, at distances from the rupture tip much larger than $x_\mathrm{c}$, the model is undistinguishable from a singular rupture model where the strength drops instantaneously to the residual value at the tip. So from a dimensional perspective, we have an intuition that the details of the cohesive zone will be masked if measurements are performed at distances $\ll x_\mathrm{c}$.

Beyond this obvious consideration, one may want to quantitatively estimate \emph{how close from the fault} a stress measurement could be deemed representative of the actual strength evolution. We can explore that using directly the results of \citet{poliakov02}. From their appendix A, we can compute the full stress field for steady-state ruptures in both the cohesive and singular case, keeping the same fracture energy in both cases.

\begin{figure}
  \centering
  \includegraphics{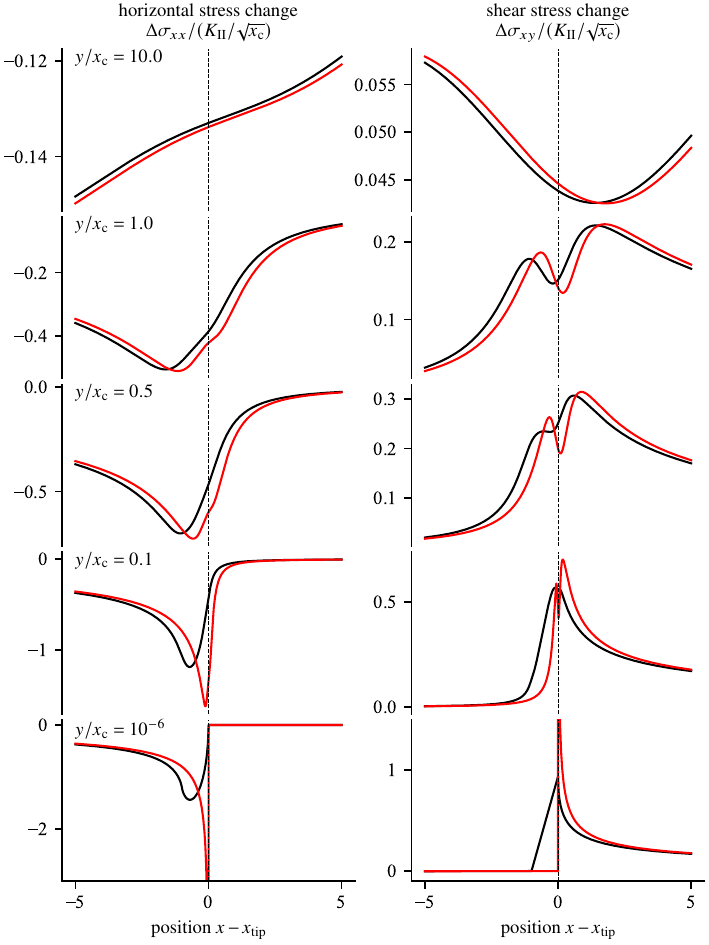}
  \caption{Horizontal and shear stress perturbations at a range of positions $y/R$ off a fault plane as a function of distance along the fault, for a steady-state rupture propagating at $C_\mathrm{f}=0.01 C_\mathrm{S}$. Red lines correspond to a singular model with instantaneous stress drop, and black lines correspond to a cohesive zone model (as depicted in Figure \ref{fig:sketch}).}
  \label{fig:stress_x_001}
\end{figure}

\begin{figure}
  \centering
  \includegraphics{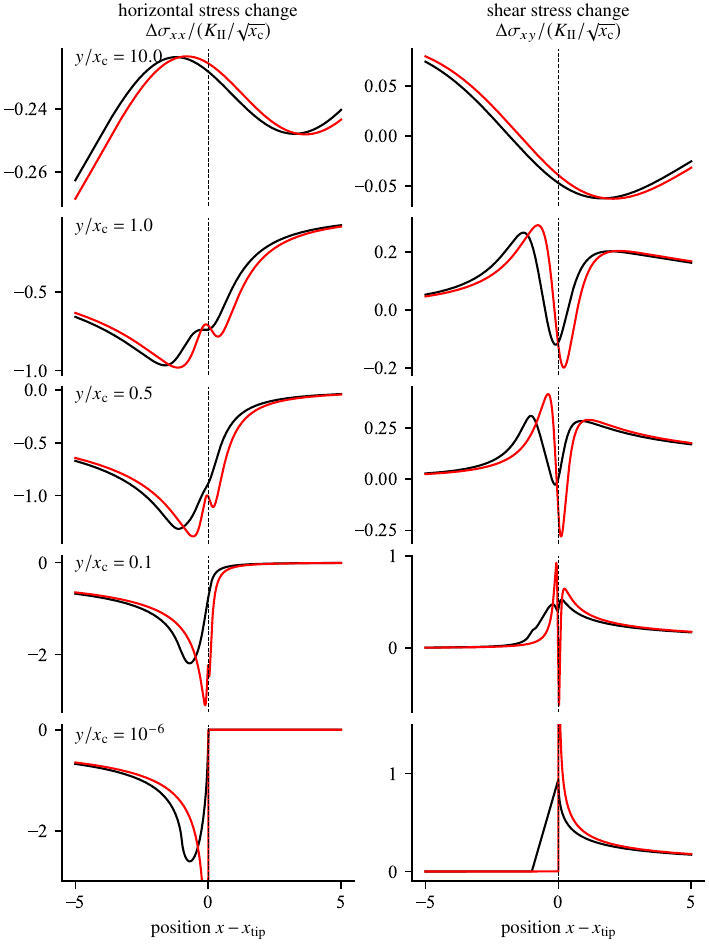}
  \caption{Horizontal and shear stress perturbations at a range of positions $y/R$ off a fault plane as a function of distance along the fault, for a steady-state rupture propagating at $C_\mathrm{f}=0.8 C_\mathrm{S}$. Red lines correspond to a singular model with instantaneous stress drop, and black lines correspond to a cohesive zone model (as depicted in Figure \ref{fig:sketch}).}
  \label{fig:stress_x_08}
\end{figure}

Fault parallel ($\sigma_{xx}$) and shear ($\sigma_{xy}$) stress profiles at different distances $y/x_\mathrm{c}$ are shown in Figures \ref{fig:stress_x_08} and \ref{fig:stress_x_001} for $C_\mathrm{f}/C_\mathrm{S}=0.8$ and $C_\mathrm{f}/C_\mathrm{S}=0.01$, respectively, where $C_\mathrm{S}$ is the shear wave velocity of the surrounding medium. For both rupture speeds, differences between singular and nonsingular stress fields are only apparent at distances $y/x_\mathrm{c}<1$, where the singular rupture displays a strong oscillation at $x=x_\mathrm{tip}$, weaker or absent for the nonsingular rupture. For $y/x_\mathrm{c}\geq1$, there are only slight quantitative differences between singular and nonsingular stress fields, the latter being offset in space due to the smeared nature of the rupture tip position, but it is clear that experimental data would have to be remarkably accurate to clearly resolve differences between the two cases, let alone provide a quantitative estimate of the cohesive zone size independently of the fracture energy. Typical horizontal strain measurements are shown in Figure 2c of \citet{berman20}, and the experimental error is clearly too large to correctly match any cohesive zone model.

An additional observation is that the shear stress evolution \emph{off the fault} is severely distorted, both in magnitude and shape, compared to that \emph{on the fault} even at distances as close as $y/x_\mathrm{c}=0.1$. Recently, \citet{kammer19} reported shear stress data from strain gauges positioned at distances of the order of $y/x_\mathrm{c}=1$ during dynamic slip events along granite fault surfaces. They showed that they could estimate, to first order, the size of the cohesive zone based on the duration and amplitude of the stress perturbation. This is a very robust observation, but it is clear that the details of the stress variations within the cohesive are not accessible from such measurements.

\section{Inversion procedure}

The shear stress evolution along the fault is approximated by a piece-wise linear function parameterised by the values of stress $\tau_i$ at positions $x_i$, where $x_0=0$ is at the rupture tip and $x_i$ ($i=1,\ldots,N$) are logarithmically spaced positions measured from the rupture tip. In all inversions, the positions $x_i$ are set to start from $x_1=1$~mm up to $x_N$ given by the maximum value of $x$ in each slip rate record.

Since we impose $\tau_N=\tau_\mathrm{b}$ (no residual stress above or below the background stress), there are $N$ unknown values of shear stress, which form the vector $\mathbf{m}$ of model parameters.

As stated in Equation (6) of the main text, the forward problem is put in the form
\begin{equation}
  \mathbf{d} = \mathbf{g}(\mathbf{m}),
\end{equation}
where $\mathrm{d}$ is the vector containing the observed slip rate data at positions $X_j$ ($j=0,\ldots,M$), and $\mathbf{g}$ is the (numerical) function producing slip rate predictions based on input stress profile. The function $\mathbf{g}$ is split in two main steps: (1) a computation of slip rate at Chebyshev nodes $z_k$ ($k=0,\ldots,N_\mathrm{k}$), and (2) an interpolation of the output onto nodes $X_j$. We set $N_\mathrm{k}=256$, keeping in mind that this choice is only important to set the overall accuracy of the forward solution method \citep{viesca18}.

The slip rate data are used ``as is'', in units of m/s. The stress values (in vector $\mathbf{m}$) are all normalised by a factor $\bar{\mu}(C_\mathrm{f})/C_\mathrm{f}$ (units of Pa/(m/s)). The positions along the fault are referenced to the rupture tip position $x=0$, which is chosen as the position where the slip rate becomes equal to or greater than the slip rate recorded at the tail-end of the rupture. The impact of that choice is not substantial, as discussed in the next Section.

The inverse problem is solved by the quasi-Newton method \citep[][Section 3.4]{tarantola05}, which minimises the misfit function
\begin{equation}
  2 S(\mathbf{m}) = \big(\mathbf{g}(\mathbf{m})-\mathbf{d}_\mathrm{obs}\big)\mathbf{C}_\mathrm{D}^{-1}\big(\mathbf{g}(\mathbf{m})-\mathbf{d}_\mathrm{obs}\big) + (\mathbf{m}-\mathbf{m}_\mathrm{prior})\mathbf{C}_\mathrm{M}^{-1}(\mathbf{m}-\mathbf{m}_\mathrm{prior}),
\end{equation}
where $\mathbf{d}_\mathrm{obs}$ is the vector of observed data, $\mathbf{C}_\mathrm{D}$ is the covariance matrix of the observed data, $\mathbf{m}_\mathrm{prior}$ is the vector of prior model parameters (also used as the initial guess for the inversion), and $\mathbf{C}_\mathrm{M}$ is the covariance matrix of the model parameters.

The covariance matrix $\mathbf{C}_\mathrm{D}$ is assumed diagonal (i.e., data points are independent of each other), and all diagonal terms are set to $1/10$, i.e., a standard deviation on each slip rate data point equal to $\sqrt{1/10}$ m/s. This specific choice has a negligible effect on the best fit solution. The elements of the covariance matrix $\mathbf{C}_\mathrm{M}$ are computed as
\begin{equation}
  (\mathbf{C}_\mathrm{M})_{ij} = \sigma_\mathrm{M}^2\exp\big(-|x_i-x_j|/\ell\big),
\end{equation}
choosing the variance $\sigma^2_\mathrm{M}=1$ and the characteristic covariance length $\ell=1$~mm. This regularisation is used to enforce a degree of smoothness of the shear stress evolution, preventing fast, unphysical oscillations between points separated by distances less than 1~mm. Without such regularisation, spurious oscillations of the inverted shear stress appear near the rupture tip, and depend on the number of points $N$ and $N_\mathrm{k}$ used in the calculation. Choosing $\ell=1$~mm removes these numerical oscillations.

The prior shear stress $\mathbf{m}_\mathrm{prior}$ is chosen to follow a mild exponential decrease in shear strength as a function of position along the fault:
\begin{equation}
  \tau_\mathrm{prior}(x) = \exp(-x/x_\mathrm{c})/6,
\end{equation}
with $x_\mathrm{c}=3$~mm. This initial choice is rather arbitrary, and its main role is to provide an initial guess that produces a slip rate profile sufficiently close to the observed ones, hence ensuring convergence of the method. Because of rather large variance $\sigma^2_\mathrm{M}$, the detailed choice of prior has no significant impact on the best fit solution, as long as convergence is observed.

The quasi-Newton method produces iterative updates for the best-fit model (in the least-square sense) following \citep[][Equation 3.93]{tarantola05}
\begin{equation}
  \mathbf{m}_{n+1} = \mathbf{m}_n -\big(\mathbf{G}_n^t\mathbf{C}_\mathrm{D}^{-1}\mathbf{G}_n + \mathbf{C}_\mathrm{M}^{-1}\big)^{-1}\big(\mathbf{G}_n^t\mathbf{C}_\mathrm{D}^{-1}(\mathbf{d} - \mathbf{d}_\mathrm{obs}) + \mathbf{C}_\mathrm{M}^{-1}(\mathbf{m}_n - \mathbf{m}_\mathrm{prior})\big),
\end{equation}
where
\begin{equation}
  \mathbf{G}_n = \left(\frac{\partial \mathbf{g}}{\partial \mathbf{m}}\right)_{\mathbf{m}_n}
\end{equation}
is the Jacobian matrix of the operator $\mathbf{g}$. The implementation of the quasi-Newton method uses automatic differentiation to compute $\mathbf{G}_n$ at each iteration \citep{revels16}.

\section{Stability of inversion results}

% changes in background stress

To determine the performance and stability of the inversion method, it is instructive to conduct inversions on a synthetic example using a two-step linear weakening in the cohesive zone, which is a realistic test case for the actual data.

\begin{figure}
  \centering
  \includegraphics{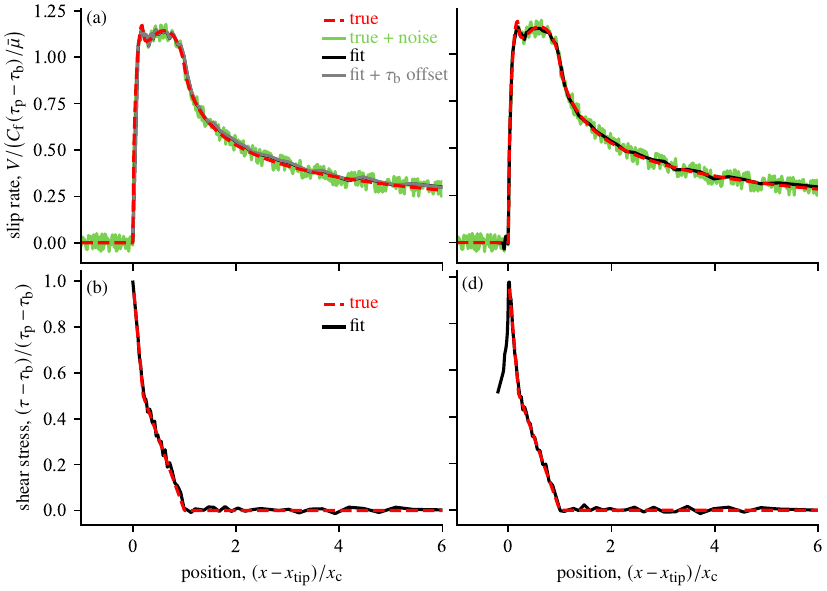}
  \caption{Synthetic tests using two-step linear stress profile. (a) Synthetic slip rate data with noise (green), data without noise (red dashed line), best-fit solution (black) and best-fit solution with added constant stress of 1 (normalised by $\tau_\mathrm{p}-\tau_\mathrm{b}$) (grey). (b) True stress profile (red dashed) and inverted stress profile (black). (c,d) same as (a,b) but offsetting rupture tip position by a $-0.2 x_\mathrm{c}$.}
  \label{fig:synthetictests}
\end{figure}

The assumed shear stress in the cohesive zone was an initial drop of 50\% from peak (denoted $\tau_\mathrm{p}$) to residual in the first 20\% of the cohesive zone size (denoted $x_\mathrm{c}$), followed by a shallower linear drop down to residual strength (equal to background stress, $\tau_\mathrm{b}$) in the remaining 80\% of the cohesive zone (Figure \ref{fig:synthetictests}b, red dashed line). The corresponding slip rate (normalised by $C_\mathrm{f}(\tau_\mathrm{p}-\tau_\mathrm{b})/\bar{\mu}$) was computed, and a random uncorrelated white noise with amplitude $\pm0.5$ added (Figure \ref{fig:synthetictests}a, green). The inverted shear stress profile (Figure \ref{fig:synthetictests}b, black line) closely matches the true solution and reproduces most features of the observed slip rate (Figure \ref{fig:synthetictests}a, black line). The added noise on the synthetic data is tracked in the solution, which contains some oscillations (especially far from the tip), but it is not detrimental to the overall shape of the stress profile in the cohesive zone.

It is instructive at this stage to observe that adding (or subtracting) a constant to the shear stress profile (e.g., to simulate a residual stress different from the background stress) has a negligible impact on the simulated slip rate (grey curve in Figure \ref{fig:synthetictests}a, obtained by offsetting the fitted stress by 1): therefore, the stress profile is only constrained to within an additive constant.

The synthetic test can also be used to assess the impact of any possible offset in the rupture tip position. With a position $x_\mathrm{tip}$ shifted to include a small region ahead of the crack, where slip rate is close to zero (within the noise level), the inverted stress is seen to include an increasing step up to the actual tip position, followed by an evolution that closely matches the true solution (Figure \ref{fig:synthetictests}d). The recovered slip rate is also close to the observed one (Figure \ref{fig:synthetictests}c), with no significant difference compared to the inversion conducted without any offset in $x_\mathrm{tip}$.

\begin{figure}
  \centering
  \includegraphics{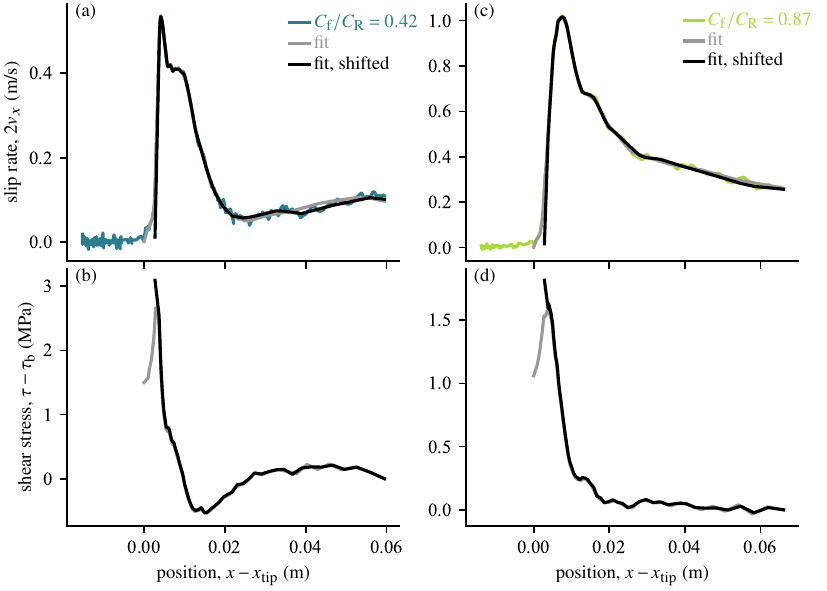}
  \caption{Comparison of inverted stress profiles with two different choices of rupture tip position in slow (a,b) and fast (c,d) ruptures. The black lines show the reference inversion cases with $x_\mathrm{tip}$ positioned where a threshold slip rate equal to that at the tail end of the event is detected. Grey lines correspond to inversions conducted with $x_\mathrm{tip}$ positioned where slip rate significantly increases above the noise level.}
  \label{fig:exampletest_xtip}
\end{figure}

This latter conclusion is reinforced by tests on actual data for both slow and fast ruptures (Figure \ref{fig:exampletest_xtip}). The stress evolution is not significantly impact by the choice of $x_\mathrm{tip}$, except for the presence of a strengthening part near the rupture tip. In practice, this strengthening is only apparent, since the accumulated slip is negligible in the region where slip rate remains small. What is effectively inverted for in that region is the (regularised) stress increase ahead of the rupture tip. %This can be seen by considering that any region ahead of the crack where llip rate is zero does not contribute to the stress redistribution in 

% changes in tip position

\bibliographystyle{agufull}
\bibliography{references}